# The dispersion of the citation distribution of top scientists' publications[1]

*Giovanni Abramo (corresponding author)*

Laboratory for Studies of Research and Technology Transfer, Institute for System Analysis and Computer Science (IASI-CNR), National Research Council of Italy
Viale Manzoni 30, 00185 Rome, Italy
giovanni.abramo@uniroma2.it

*Ciriaco Andrea D'Angelo*

Department of Engineering and Management, University of Rome 'Tor Vergata' and Laboratory for Studies of Research and Technology Transfer, Institute for System Analysis and Computer Science (IASI-CNR)
Via del Politecnico 1, 00133 Rome, Italy
dangelo@dii.uniroma2.it

*Anastasiia Soldatenkova*

Department of Engineering and Management, University of Rome 'Tor Vergata'
Via del Politecnico 1, 00133 Rome, Italy
anastasiia.soldatenkova@uniroma2.it

**Abstract**

This work explores the distribution of citations for the publications of top scientists. A first objective is to find out whether the 80-20 Pareto rule applies, that is if 80% of the citations to a top scientist's work concern 20% of their publications. Observing that the rule does not apply, we also measure the dispersion of the citation distribution by means of the Gini coefficient. Further, we investigate the question of what share of a top scientist' publications go uncited. Finally, we study the relation between the dispersion of the citation distribution and the share of uncited publications. As well as the overall level, the analyses are carried out at the field and discipline level, to assess differences across them.

---

# 1. Introduction

One of the aims of universities and research institutions is to contribute to socio-economic development through the production of new knowledge. For this the knowledge must be disseminated, and therefore codified. In the sciences, the preferred mode of codification is publication in scientific journals. Some of the publications, meaning the knowledge they embed, will have more impact than others on scientific advancement. To measure the impact, bibliometricians use the proxy of the number of citations received. For publications of the same year and field, the higher is the number of citations then the higher is the impact of scientific advancement. Uncited publications are held to have no impact at all. Further, it is held that researchers who produce higher total impact over a period of time, all other factors equal, are more productive than others.

The area of interest for the current paper is the question of the impact-output relation in research activities. We can expect that a researcher's publication portfolio will be made up of publications of different impact. A first question is whether the Pareto principle, also known as the 80-20 rule, holds true. In other words, we want to find out whether roughly 80% of the scientist's total impact comes from 20% of their publications. The answer will vary among researchers, and likely across fields and disciplines.

A more general way to state the objective of the work is that we intend to analyze the dispersion of the citation distribution of researchers' publications. For this, the Gini coefficient is a particularly useful instrument.

The analysis will focus on top scientists, because to have meaningful results requires the observation of researchers with very high numbers of articles and/or total citations. A top scientist is generally defined as a researcher whose impact on scientific advancement is in the top X% among colleagues in the same field. In this study we observe the top 10% of scientists. A top 10% scientist results as such, because: i) they produce a very high number of articles, mostly moderately cited; or ii) they publish only few articles, but highly-cited ones; or iii) something in between. Gupta, Campanha, and Pesce, (2005) state that "it is not possible for anyone to compete both in quality and quantity. Some people may have a large number of articles but a small average citation per article, while other people may have larger average citation of an article but small number of articles". Based on our experience in the assessment of research performance by Italian scientists at the individual level (Abramo & D'Angelo, 2011), we would hazard to state that the top 1% scientists in their respective fields are likely to produce very high numbers of articles, mostly highly cited. The average researcher, who contributes fewer, lesser cited articles, is not examined in the current analysis.

Scientometricians have carried out Pareto analyses in the past. The literature presents quite a number of studies at the journal level, meaning the application of the 80-20 rule concerning the articles and relative citations of a journal (Weingart, 2004; Glanzel & Moed, 2002; Seglen, 1997; Moed & Van Leeuwen, 1996). Only a few studies analyze the share of scientists who contribute the most to the total publications or the impact achieved by an entire research system. Flegl and Vydrova (2014) showed that the Pareto rule does not fit concerning the publications of PhD students at the Czech University of Life Sciences Prague, where a very large portion of the students result as totally unproductive during their PhD program. Abramo, Cicero, and D'Angelo, (2011) found that 23% of the professors in the sciences counted for 77% of the total impact of Italian universities' research over the 2004-2008 period, which is the same period that we observe in the current work. Other studies have been concerned with the statistical distribution of citations, as well as the trade-off between quantity and quality in research output (Piro, Rørstad K., & Aksnes, 2016; Bosquet & Combes, 2013; Parker, Allesina, & Lortie, 2013; Abramo, D'Angelo, & Di Costa, 2010; Gupta, Campanha, & Pesce, 2005). Perianes-Rodriguez and Ruiz-Castillo (2015) investigate the citation distributions at university level. Ruiz-Castillo and Costas (2014) measure the individual performance in two ways, one being the mean citation per article per person. They then analyze the statistical distributions.

However, we find no analyses of the Pareto distribution of citations per publication at the individual level. This is probably due to the formidable problem of large-scale address reconciliation and authors' name disambiguation. For Italian publications indexed in the WoS, the problem has been overcome by means of a disambiguation algorithm developed by D'Angelo, Giuffrida, and Abramo, (2011). Using the algorithm, we can construct the publication portfolios of all Italian professors over any period of time, and produce performance rankings by any bibliometric indicator at the individual level, on a national scale. We can thus identify all top scientists (TSs) in Italian universities in the different fields of research in the sciences, and then analyze the statistical dispersion of the distribution of citations for their publications, including the question of the Pareto distribution. An equally interesting inquiry about the output of TSs could be to examine the distribution of the share of papers that remains uncited. In the current paper we carry out this analysis. Finally, for each professor, we will explore the correlation between the minimum number of publications which account for at least 80% of total citations, and the percentage of papers that are uncited.

The following section presents the data and method. In Section 3 we report the results of the analyses, which we discuss in the concluding section.

## 2. Data and methods

The field of observation is composed of the 2004-2008 publications of Italian academic TSs in the sciences, as indexed in the WoS. The citations are counted for all the publications as of June 2015, with the intention that the seven-year and up citation window ensures a robust measure of impact.

In the Italian academic system, each professor is classified in one and only one research field. There are a total of 370 such fields (named "scientific disciplinary sectors", or SDSs[2]), grouped into 14 disciplines (named "university disciplinary areas", or UDAs). The Italian Ministry of University and Research (MIUR) maintains a database of all Italian professors (http://cercauniversita.cineca.it). For each individual, the MIUR database shows the last name and given names, university, SDS, academic rank and department. Our analysis is limited to the professors working in the nine science UDAs (192 SDSs) where scientific performance can be assessed using bibliometric techniques with an acceptable level of reliability: Mathematics and computer science, Physics, Chemistry, Earth sciences, Biology, Medicine, Agricultural and veterinary sciences, Civil engineering, and Industrial and information engineering.

The bibliometric data of each professor is extracted from the Italian Observatory of

---

[2] The complete list is accessible at http://attiministeriali.miur.it/UserFiles/115.htm. Last accessed on September 6, 2016.

Public Research (ORP), a database developed and maintained by the authors and derived under license from Thomson Reuters WoS. Beginning from the raw data of the WoS, and applying the algorithm to reconcile the authors' affiliation and disambiguate their identity, each publication (article, article review, letter and conference proceeding) is attributed to the university professors(s) that produced it.[3]

To identify the TSs in each SDS, we rank all professors by their total impact in the period under observation. Because citation behavior varies by field, we field normalize the citations. The ranking is developed by measuring the yearly average total impact (Scientific Strength or SS) of each professor, summing up the field-normalized impact of all their publications indexed in the period under observation. In formulae:

$$SS = \frac{1}{t}\sum_{i=1}^{N}\frac{c_i}{\bar{c}}$$

Where:
$t$ = number of years of work of the professor in period under observation[4]
$N$ = number of publications of the professor in period under observation
$c_i$ = citations received by publication $i$
$\bar{c}$ = average of distribution of citations received for all cited Italian publications in same year and subject category of publication $i$[5]

The dataset for the analysis (Table 1) is then made up of the top 10% professors by SS in each SDS. To ensure the robustness of the statistics calculated, any SDS with less than 10 TSs is excluded from the analysis. Thus organized, the dataset is composed of 75,184 publications authored or co-authored by a total of 3,386 TSs in Italian universities, sorted in 9 UDAs comprising a total of 130 SDSs.

*Table 1: Dataset for the analyses: number of SDSs, top scientists (TSs) and publications in each UDA (2004-2008 data)*

| UDA | N. of SDSs | Top scientists | TSs Publications |
|---|---|---|---|
| 1 - Mathematics and computer science | 9 | 329 | 5,519 |
| 2 - Physics | 6 | 252 | 6,614 |
| 3 - Chemistry | 8 | 312 | 10,349 |
| 4 - Earth sciences | 8 | 108 | 1,657 |
| 5 - Biology | 17 | 507 | 12,286 |
| 6 - Medicine | 37 | 1,076 | 30,315 |
| 7 - Agricultural and veterinary sciences | 15 | 230 | 3,303 |
| 8 - Civil engineering | 8 | 149 | 1,839 |
| 9 - Industrial and information engineering | 22 | 423 | 10,043 |
| Total | 130 | 3,386 | 75,184 |

**3. Results**

We analyze the citation distribution of top scientists' publications using two measures of inequality. First, we verify if the contribution of publications to total impact follows the Pareto rule: i.e. whether 80% of the citations come from 20% of the publications. In a second part, we measure the statistical dispersion of the distribution of citations by means of the Gini coefficient. Finally, we will analyze the percentage of TSs' publications with nil citations, at the individual level. All the analyses are carried out at the overall level, as well as by SDS and UDA, to assess differences across fields and disciplines.

---

[3] The harmonic mean of precision and recall (F-measure) of authorships disambiguated by the algorithm is around 97% (2% margin of error, 98% confidence interval).
[4] Professors with less than three years on duty were excluded from the analysis.
[5] Abramo, Cicero, and D'Angelo, (2012) demonstrated that this is the best-performing scaling factor.

**3.1 Pareto rule**

For each TS, we calculate the minimum percentage of publications indexed in the period under observation which account for at least 80% of total field-normalized citations (MPP). As an example, in Figure 1 we show the Pareto chart concerning a professor classified in Automatics (SDS: ING-INF/04), who produced 146 WoS-indexed publications in the period under examination. In this case we can observe a proportion close to the 80-20 rule: 21% of his publications (31 out of 146) account for 80.7% of the total field-normalized citations.[6] The maximum number of citations received by a single paper equals 5.5, and the distribution decreases quite evenly. The long right tail represents 63 (43%) uncited publications.

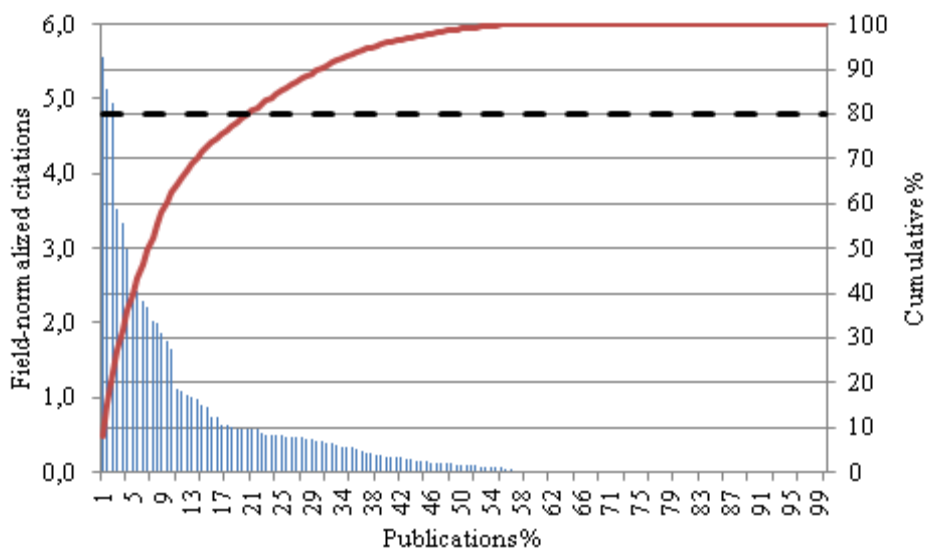

*Figure 1: Pareto chart of a top scientist in Automatics (data 2004-2008)*

Figure 2 shows the MPP frequency distribution of all TSs. The highest frequencies occur in the range 35% - 55%, accounting for 62% of TSs (2,113 cases out of 3,386). The mode is 50%. There are no cases above 85% and below 100%. The last bar represents the 100% MPP value, which occurs in 30 cases (0.89%). The minimum MPP value is instead 3%, occurring in the sole case of a TS in Computer science (INF/01).

In Table 2 we report the descriptive statistics of the MPP distribution.

We now proceed with the analysis at the SDS level, to see if noticeable differences occur in the MPP frequency distributions across SDSs. Table 3 shows the descriptive statistics of the MPP distribution for the SDSs within Chemistry (UDA 3). The maximum MPP value varies between 50.0% and 62.5%, while the minimum shows greater differences, from 5.0% to 33.3%. The lowest standard deviation (5.7%) occurs in Foundations of chemistry for technologies (CHIM/07).

---
[6] For ease of presentation, in the following we use "citations" in place of "field-normalized citations".

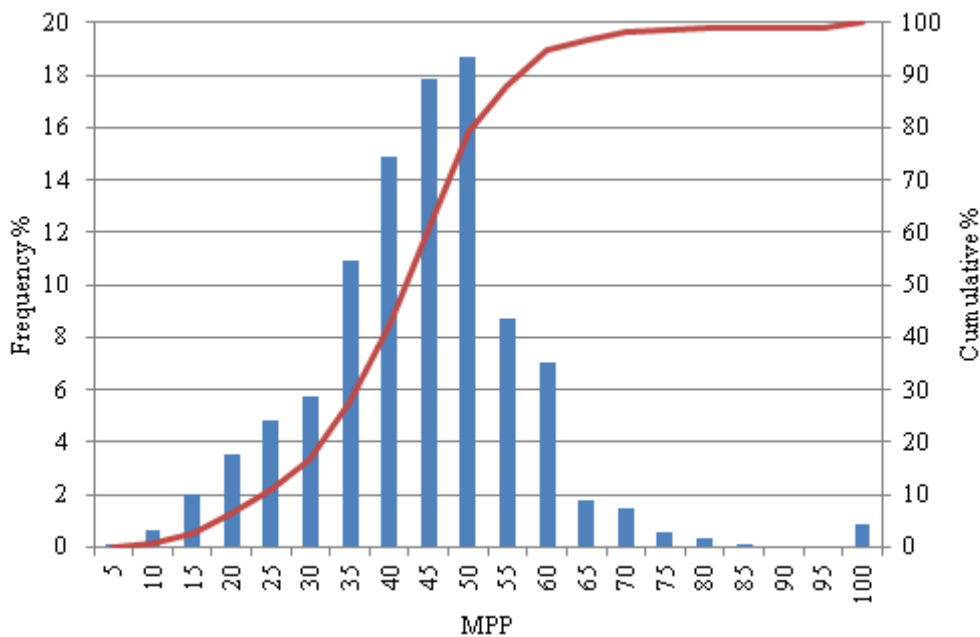

*Figure 2: MPP frequency distribution for all Italian top scientists in the sciences (2004-2008 data)*

*Table 2: Descriptive statistics of the MPP distribution*

| | |
|---|---|
| Mean | 0.42 |
| Median | 0.43 |
| Mode | 0.50 |
| Standard Deviation | 0.13 |
| Sample Variance | 0.02 |
| Kurtosis | 2.39 |
| Skewness | 0.43 |
| Range | 0.97 |
| Minimum | 0.03 |
| Maximum | 1 |
| Count | 3386 |

*Table 3: Descriptive statistics of the MPP distribution in the SDSs of Chemistry*

| SDS* | N. of TSs | MPP | | | |
|---|---|---|---|---|---|
| | | Average | Min | Max | std dev. |
| CHIM/01 | 29 | 43.5% | 21.7% | 55.6% | 9.0% |
| CHIM/02 | 48 | 40.7% | 6.3% | 53.8% | 12.1% |
| CHIM/03 | 62 | 44.9% | 5.0% | 54.5% | 9.3% |
| CHIM/04 | 16 | 43.1% | 30.0% | 51.5% | 7.8% |
| CHIM/06 | 68 | 43.0% | 15.0% | 57.1% | 10.1% |
| CHIM/07 | 20 | 43.2% | 33.3% | 50.0% | 5.7% |
| CHIM/08 | 48 | 48.5% | 26.7% | 62.5% | 8.5% |
| CHIM/09 | 21 | 44.1% | 15.4% | 54.5% | 11.7% |

*\* CHIM/01=Analytical Chemistry; CHIM/02=Physical Chemistry; CHIM/03=General and Inorganic Chemistry; CHIM/04=Industrial Chemistry; CHIM/05=Science and Technology of Polymeric Materials; CHIM/06=Organic chemistry; CHIM/07=Foundations of Chemistry for Technologies; CHIM/08=Pharmaceutical Chemistry; CHIM/09=Applied Technological Pharmaceutics;*

*CHIM/10=Food Chemistry; CHIM/11=Chemistry and Biotechnology of Fermentations; CHIM/12=Environmental Chemistry and Chemistry for Cultural Heritage*

Table 4 presents the SDSs in each UDA with the minimum and maximum average values of MPP. The SDS with the minimum average MPP (19.1%) is Electrical convertors, machines and switches (ING-IND/32). This is the only SDS for which the Pareto rule applies. The SDS with the maximum value (71.1%) is Road, railway and airport construction (ICAR/04). The lowest difference between the min-max values (7.8%) occurs in Chemistry (UDA 3), while the largest one (47.0%) occurs in Industrial and information engineering (UDA 9),

*Table 4: SDSs\* with min/max average MPP in each UDA*

| UDA | N. of SDSs | Average MPP | | | | |
|---|---|---|---|---|---|---|
| | | Min | | Max | | Δ |
| 1 | 9 | INF/01 | 26.0% | MAT/02 | 52.9% | 26.9% |
| 2 | 6 | FIS/04 | 25.5% | FIS/05 | 40.0% | 14.5% |
| 3 | 8 | CHIM/02 | 40.7% | CHIM/08 | 48.5% | 7.8% |
| 4 | 8 | GEO/06 | 38.1% | GEO/01 | 57.2% | 19.1% |
| 5 | 17 | BIO/18 | 39.9% | BIO/03 | 54.9% | 15.0% |
| 6 | 37 | MED/06 | 30.9% | MED/43 | 57.5% | 26.6% |
| 7 | 15 | AGR/19 | 36.6% | AGR/13 | 59.2% | 22.6% |
| 8 | 8 | ICAR/07 | 40.1% | ICAR/04 | 71.1% | 30.9% |
| 9 | 22 | ING-IND/32 | 19.1% | ING-IND/35 | 66.1% | 47.0% |

\* *INF/01=Computer Science; MAT/02=Algebra; FIS/04=Nuclear and Subnuclear Physics; FIS/05=Astronomy and Astrophysics; CHIM/02=Physical Chemistry; CHIM/08=Pharmaceutical Chemistry; GEO/06=Mineralogy; GEO/01=Palaeontology and Palaeoecology; BIO/18=Genetics; BIO/03=Environmental and Applied Botanics; MED/06=Medical Oncology; MED/43=Legal Medicine; AGR/19=Special Techniques for Zoology; AGR/13=Agricultural Chemistry; ICAR/07=Geotechnics; ICAR/04=Road, Railway and Airport Construction; ING-IND/32= Electrical Convertors, Machines and Switches; ING-IND/35=Engineering and Management*

An interesting focus concerns the analysis of the variability of MPP values within and between SDSs. In this regard, Table 5 shows, for each UDA, the SDSs with minimum and maximum variation coefficient of MPP. It also shows the variability between SDSs, i.e. the variation coefficient of the distribution of average MPP of SDSs of each UDA. We note that the variability between is systematically lower than that within SDSs, with the only exception of 10 SDSs (listed in the last column of Table 5) out of 130. In short, only in Industrial and information engineering there is a significant number of SDSs (7 out of 22) whereby the MPP variability between is higher than within.

*Table 5: SDSs\* with min/max MPP variation coefficient in each UDA and variation coefficient of MPP between SDSs of the UDA*

|     | Within (min variability) | | | Within (max variability) | | | Between SDSs | | |
|-----|------|---------|-------------|------|---------|-------------|-------------|-------------|-------------|
| ADU | SDS | Obs (TS) | Var. coeff. | SDS | Obs (TS) | Var. coeff. | Obs (SDSs) | Var. coeff. | No. of SDSs with MPP variation coefficient lower than "between" SDS |
| 1 | MAT/08 | 25 | 0.236 | INF/01 | 71 | 0.481 | 9 | 0.172 | 0 |
| 2 | FIS/05 | 18 | 0.187 | FIS/04 | 17 | 0.389 | 6 | 0.152 | 0 |
| 3 | CHIM/07 | 20 | 0.135 | CHIM/02 | 48 | 0.300 | 8 | 0.051 | 0 |
| 4 | GEO/03 | 11 | 0.145 | GEO/06 | 11 | 0.360 | 8 | 0.123 | 0 |
| 5 | BIO/04 | 12 | 0.086 | BIO/18 | 21 | 0.351 | 17 | 0.081 | 0 |
| 6 | MED/15 | 18 | 0.117 | MED/10 | 13 | 0.406 | 37 | 0.127 | 1 (MED/15) |
| 7 | VET/05 | 11 | 0.118 | AGR/01 | 34 | 0.437 | 15 | 0.117 | 0 |
| 8 | ICAR/01 | 14 | 0.144 | ICAR/05 | 10 | 0.507 | 8 | 0.199 | 2 (ICAR/01,ICAR/08) |
| 9 | ING-IND/22 | 24 | 0.108 | ING-IND/08 | 16 | 0.533 | 22 | 0.299 | 7 (ING-IND/11, ING-IND/14, ING-IND/16, ING-IND/22, ING-IND/25,ING-INF/06, ING-INF/07) |

\* MAT/08=Numerical analysis; FIS/05=Astronomy and Astrophysics; CHIM/07=Foundations of Chemistry for Technologies; GEO/03=Structural Geology; BIO/04=Vegetal Physiology; MED/15=Blood Diseases; VET/05=Infectious Diseases of Domestic Animals; ICAR/01=Hydraulics; ING-IND/22=Science and Technology of Materials; INF/01=Computer Science; FIS/04=Nuclear and Subnuclear Physics; CHIM/02=Physical Chemistry; GEO/06=Mineralogy; BIO/18=Genetics; MED/10=Respiratory Diseases; AGR/01=Rural economy and evaluation; ICAR/05=Transport; ING-IND/08=Fluid Machines; ICAR/08=Construction Science; ING-IND/11=Environmental Technical Physics; ING-IND/14=Mechanics and Machine Design; ING-IND/16=Production Technologies and Systems; ING-IND/25=Chemical Plants; ING-INF/06=Electronic and Information Bioengineering; ING-INF/07=Electric and Electronic Measurement Systems

## 3.2 Gini coefficient

Another way to investigate and report the inequality in the distribution of citations for the publications of each TS is by means of the Gini coefficient of the distribution. The value 0 expresses perfect equality, in which each publication accumulates an equal amount of year- and field-standardized citations. The closer the value is to 1, the stronger is the degree of inequality.

As in the previous subsection, we compute the Gini coefficient for each TS in the dataset. We then aggregate the results at the SDS and UDA levels. Figure 3 shows the box plot of the average values of Gini coefficient as recorded for each of the 130 SDSs under consideration. The box represents the inter-quartile range (IQR), meaning the interval of observations between the first (0.48) and third quartile (0.58) (respectively indicated by the lower and upper borders of the box). The line that divides the box in two parts represents the median (0.53). The whiskers below and above the box represent the lowest observation still within 1.5 IQR of the lower quartile, and the highest observation still within 1.5 IQR of the upper quartile. The graph also indicates three outliers, marked by dots. The first one equals 0.79 and refers to the disciplinary sector of Electrical convertors, machines and switches (ING-IND/32), the other two at the bottom, equal respectively 0.30 (Road, railway and airport construction, ICAR/04) and 0.32 (Engineering and management, ING-IND/35).

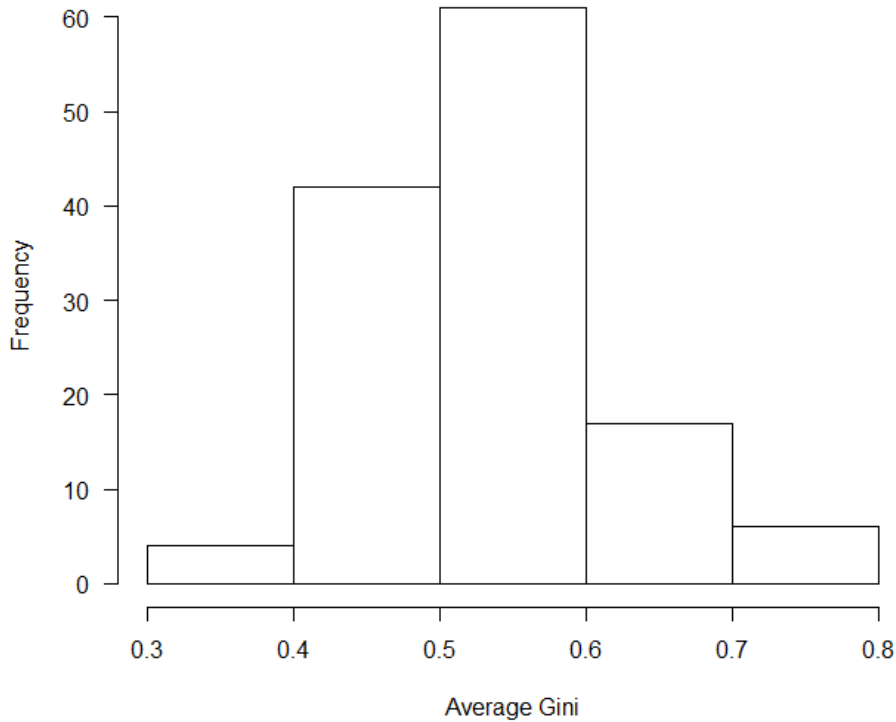

*Figure 3: Distribution of average values of Gini coefficients for TSs in the 130 SDSs under investigation*

Table 6 presents the SDSs within each UDA that show the minimum and maximum average value of Gini coefficient. The results are similar to those shown in Table 4. The SDSs that showed a minimum average value of MPP obviously now show maximum values of Gini coefficient, and vice versa. However there are three exceptions, in Mathematics and computer science (UDA 1), Earth sciences (UDA 4), and Agricultural and veterinary sciences (UDA 7), where the SDSs with the maximum value of average MPP are different from those with minimum of Gini coefficient.

*Table 6: SDSs\* with min/max average Gini coefficient in each UDA*

| UDA | N. of SDSs | Average Gini | | | | | |
|---|---|---|---|---|---|---|---|
| | | Min | | Max | | Δ | |
| 1 | 9 | MAT/04 | 0.441 | INF/01 | 0.714 | 0.273 | |
| 2 | 6 | FIS/05 | 0.565 | FIS/04 | 0.708 | 0.143 | |
| 3 | 8 | CHIM/08 | 0.473 | CHIM/02 | 0.562 | 0.089 | |
| 4 | 8 | GEO/02 | 0.400 | GEO/06 | 0.577 | 0.178 | |
| 5 | 17 | BIO/03 | 0.399 | BIO/18 | 0.567 | 0.168 | |
| 6 | 37 | MED/43 | 0.406 | MED/06 | 0.664 | 0.258 | |
| 7 | 15 | AGR/01 | 0.412 | AGR/19 | 0.593 | 0.181 | |
| 8 | 8 | ICAR/04 | 0.304 | ICAR/07 | 0.592 | 0.289 | |
| 9 | 22 | ING-IND/35 | 0.315 | ING-IND/32 | 0.786 | 0.471 | |

\* INF/01=Computer Science; MAT/04=Complementary Mathematics; FIS/04=Nuclear and Subnuclear Physics; FIS/05=Astronomy and Astrophysics; CHIM/02=Physical Chemistry; CHIM/08=Pharmaceutical Chemistry; GEO/06=Mineralogy; GEO/02=Stratigraphic and Sedimentological Geology; BIO/18=Genetics; BIO/03=Environmental and Applied Botanics; MED/06=Medical Oncology; MED/43=Legal Medicine; AGR/19=Special Techniques for Zoology; AGR/01=Rural economy and evaluation; ICAR/07=Geotechnics; ICAR/04=Road, Railway and Airport Construction; ING-IND/32= Electrical Convertors, Machines and Switches; ING-IND/35=Engineering and Management

### 3.3 The percentage of uncited publications out of total

When we think of top scientists, we imagine researchers with outstanding performance, who produce high numbers of publications that also receive high numbers of citations. We would be surprised if a noticeable share of a TS's publication portfolio were to go uncited. The focus of this subsection is indeed the uncited side of the TSs' research output.

For each TS in the dataset, we calculate the percentage of uncited publications (PUP) out of their total publications in the period under observation. Figure 4 shows the overall frequency distribution of PUP. The minimum value of 0% occurs in 1,183 out of 3,386 (35%) cases. The maximum value (71.4%) occurs for a TS in Electrical energy systems (ING-IND/33), for whom 15 of 21 publications went uncited. Overall, the greatest number of cases (1,876 out of 3,386, or 55.4%) are concentrated in the range of 0% to 5%.

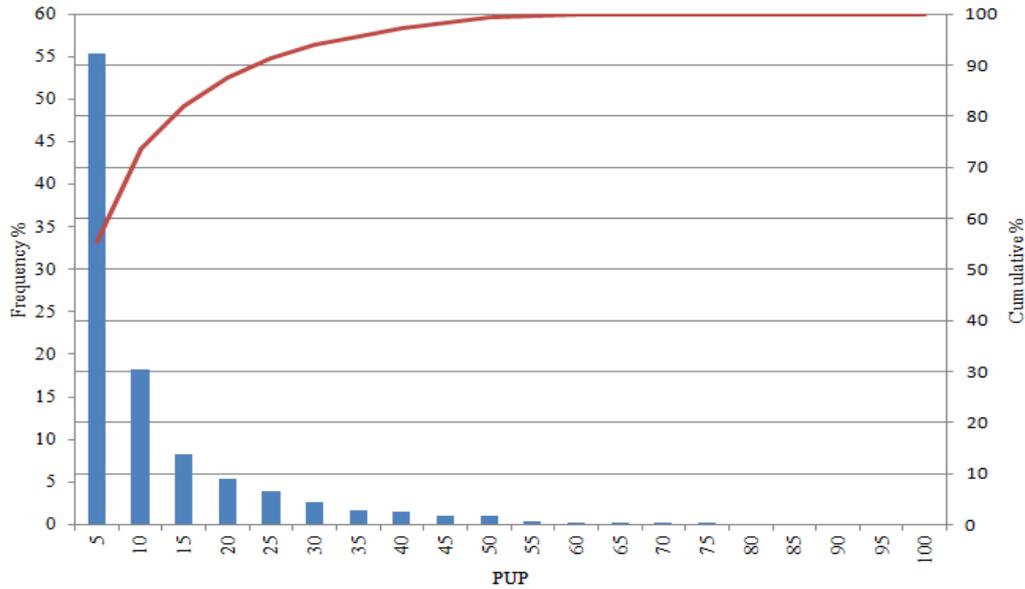

*Figure 4: Frequency distribution of percentage of uncited publications (PUP) of all TSs in the dataset*

We also calculate the descriptive statistics for the PUP values at the SDS level. As an example, Table 7 presents the statistics for the SDSs in Industrial and information engineering (UDA 9). This is the only UDA where we observe SDSs with minimum values of PUP (column 4) other than 0: here there are 11 such SDSs out of the total 22 (50%). This is probably due to the high incidence of conference proceedings in this discipline, along with the behavior of citing them less than other document types. Within the UDA we observe two extreme values for the overall distributions, occurring in the two SDSs with the highest average PUP: the highest minimum value, of 23.9%, is in Electrical convertors, machines and switches (ING-IND/32), while the highest maximum value, of 71.4%, occurs in Electrical energy systems (ING-IND/33).

To examine the differences between the nine UDAs, Table 8 presents the SDSs with the minimum and maximum average values of PUP, in each UDA. An interesting case is Palaeontology and palaeoecology GEO/01 (UDA 4), where all TSs had all publications cited. We see that UDA 9 is the one with the greatest difference between the minimum and maximum average PUP values for its SDSs (33.2%). This is also the UDA where we observe the SDSs with the highest max (39.3%) and the highest min (6.2%) of average PUP.

*Table 7: Descriptive statistics of percentage of uncited publications (PUP) for TSs in the SDSs of Industrial and information engineering (UDA 9)*

| SDS | N. of TSs | PUP (%) | | | |
|---|---|---|---|---|---|
| | | Average | Min | Max | std dev |
| ING-IND/08 | 16 | 18.0 | 0.0 | 50.0 | 15.8 |
| ING-IND/10 | 17 | 22.8 | 0.0 | 53.8 | 17.7 |
| ING-IND/11 | 19 | 17.2 | 0.0 | 33.3 | 10.5 |
| ING-IND/13 | 19 | 31.2 | 0.0 | 53.3 | 17.0 |
| ING-IND/14 | 18 | 12.9 | 0.0 | 25.0 | 8.2 |
| ING-IND/15 | 10 | 34.5 | 12.5 | 57.1 | 13.1 |
| ING-IND/16 | 16 | 17.6 | 4.4 | 40.0 | 10.6 |
| ING-IND/17 | 13 | 11.1 | 0.0 | 26.7 | 9.1 |
| ING-IND/21 | 10 | 14.6 | 4.3 | 40.9 | 9.8 |
| ING-IND/22 | 24 | 6.7 | 0.0 | 14.0 | 4.1 |
| ING-IND/25 | 11 | 7.2 | 0.0 | 16.7 | 4.7 |
| ING-IND/31 | 20 | 31.8 | 2.3 | 66.7 | 17.0 |
| ING-IND/32 | 12 | 39.3 | 23.9 | 48.5 | 8.4 |
| ING-IND/33 | 12 | 35.3 | 10.0 | 71.4 | 17.7 |
| ING-IND/35 | 16 | 6.2 | 0.0 | 21.1 | 8.5 |
| ING-INF/01 | 35 | 21.8 | 5.3 | 42.1 | 8.2 |
| ING-INF/02 | 17 | 26.3 | 4.7 | 55.0 | 14.0 |
| ING-INF/03 | 30 | 34.8 | 10.7 | 61.2 | 11.5 |
| ING-INF/04 | 26 | 25.9 | 7.7 | 43.2 | 10.2 |
| ING-INF/05 | 60 | 27.4 | 0.0 | 60.0 | 12.5 |
| ING-INF/06 | 10 | 16.8 | 0.0 | 42.9 | 12.6 |
| ING-INF/07 | 12 | 34.6 | 17.4 | 47.6 | 8.9 |

\* ING-IND/08=Fluid Machines; ING-IND/10=Technical Physics; ING-IND/11=Environmental Technical Physics; ING-IND/13=Applied Mechanics for Machinery; ING-IND/14=Mechanics and Machine Design; ING-IND/15=Design and Methods for Industrial Engineering; ING-IND/16=Production Technologies and Systems; ING-IND/17=Industrial and Mechanical Plant; ING-IND/21=Metallurgy; ING-IND/22=Science and Technology of Materials; ING-IND/25=Chemical Plants; ING-IND/31=Electrotechnics; ING-IND/32=Electrical Convertors, Machines and Switches; ING-IND/33=Electrical Energy Systems; ING-IND/35=Engineering and Management; ING-INF/01=Electronics; ING-INF/02=Electromagnetic Fields; ING-INF/03=Telecommunications; ING-INF/04=Automatics; ING-INF/05=Data Processing Systems; ING-INF/06=Electronic and Information Bioengineering; ING-INF/07=Electric and Electronic Measurement Systems

*Table 8: SDSs with min/max average percentage of uncited publications (PUP) in each UDA*

| UDA | N. of SDSs | Average PUP | | | | Δ |
|---|---|---|---|---|---|---|
| | | Min | | Max | | |
| 1 | 9 | MAT/06 | 5.4 | INF/01 | 25.8 | 20.4 |
| 2 | 6 | FIS/01 | 5.0 | FIS/07 | 10.0 | 5.0 |
| 3 | 8 | CHIM/08 | 2.3 | CHIM/07 | 6.7 | 4.4 |
| 4 | 8 | GEO/01 | 0.0 | GEO/05 | 10.5 | 10.5 |
| 5 | 17 | BIO/11 | 1.0 | BIO/18 | 4.1 | 3.1 |
| 6 | 37 | MED/05 | 1.8 | MED/23 | 11.6 | 9.8 |
| 7 | 15 | AGR/11 | 1.9 | AGR/19 | 14.1 | 12.2 |
| 8 | 8 | ICAR/06 | 3.4 | ICAR/07 | 25.7 | 22.4 |
| 9 | 22 | ING-IND/35 | 6.2 | ING-IND/32 | 39.3 | 33.2 |

\* INF/01=Computer Science; MAT/06=Probability and Mathematical Statistics; FIS/07=Applied Physics (Cultural Heritage, Environment, Biology and Medicine); FIS/01=Experimental Physics; CHIM/07=Foundations of Chemistry for Technologies; CHIM/08=Pharmaceutical Chemistry; GEO/05=Applied Geology; GEO/01=Palaeontology and Palaeoecology; BIO/18=Genetics; BIO/11=Molecular Biology; MED/23=Cardiac Surgery; MED/05=Clinical Pathology; AGR/19=Special Techniques for Zoology; AGR/11=General and Applied Entomology; ICAR/07=Geotechnics; ICAR/06=Topography and Cartography; ING-IND/32= Electrical Convertors, Machines and Switches; ING-IND/35=Engineering and Management

**3.4 Correlation between the MPP and PUP distributions**

The TSs whose citation distribution is highly dispersed (high MPP and low Gini coefficient) are likely to present a low percentage of uncited publications (low PUP). Vice versa, TSs with highly concentrated citation distribution are more likely to present a high percentage of uncited publications. The scatter plot of Figure 5 positions each TS in terms of MPP and PUP. To verify these expectations we calculate the Pearson correlation coefficient ($\rho = -0.50$), which reveals only a weak relationship between MPP and PUP. The R-squared value equals 0.25.

A further question that interests us is whether the TSs with a higher number of publications might also present a higher number of uncited publications. The correlation analysis reveals no correlation at all between the two variables ($\rho = -0.07$).

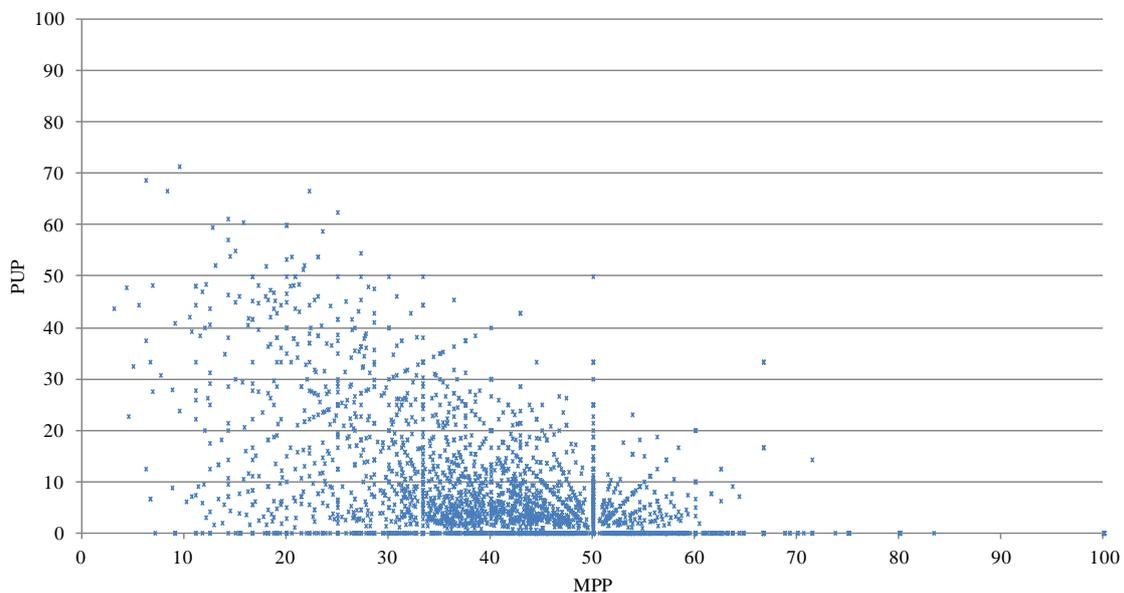

*Figure 5: Scatter plot of MPP and PUP of Italian top scientists*

**4. Conclusions**

The Pareto rule does not apply to the citation distribution of TSs' publications. For only 6% of TSs, 20% or less of their publications account for at least 80% of their total citations. A full 62% of TSs require from 35% to 55% of their total publications to reach at least 80% of their total citations. The highest TS frequency occurs at 50%, meaning that in most cases the TSs require a minimum of 50% of their publications to reach at least 80% of their total citations. The mirror view of the citation distribution, by the Gini coefficient, in fact shows that the dispersion is at the middle, between none and total. Differences occur among disciplines and among fields in the same discipline. Sectorial (SDS) differences are most noticeable when comparing the minimum/maximum values of the average minimum percentage of publications accounting for at least 80% of total citations, as well as in comparing the Gini coefficients.

Among Italian TSs, 35% have all their publications cited, and 55% have a maximum of 5% of their publications uncited. Concerning the uncited articles, differences again occur across fields and disciplines. However, we found only a weak correlation between the concentration of the citation distribution and the share of a TS's articles that go uncited. Further, there is no correlation at all between the intensity of publication and the share of uncited articles.

We can conclude that the scientists who contribute the most to scientific advancement, i.e. top scientists, in general do so through around half of their total research output. From another perspective, we can state that in the pursuit of scientific advancement by top scientists, a substantial part of their production efforts will not pay off in terms of impact.